\documentclass{emulateapj}

\usepackage{amsmath,amssymb}
\usepackage{natbib}
\usepackage{epsfig,graphics}
\usepackage{epstopdf} 
\usepackage{ulem}

\begin{document}

\slugcomment{Draft Version; not for circulation}
\title{No evidence for additional planets at GJ 3470 from TESS and archival radial velocities}

\author{Thomas W. Tarrants\altaffilmark{1}}
\author{Andrew Li\altaffilmark{2}}
\altaffiltext{1}{Corresponding Email: twtarrants@gmail.com}
\altaffiltext{2}{Corresponding Email: andrewli6907@gmail.com}

\begin{abstract}
The nearby M2 dwarf GJ 3470 has been the target of considerable interest after the discovery of a transiting short-period Neptune-sized planet. Recently, claims regarding the existence of additional transiting planets has gotten some attention, suggesting both the presence of a gas giant in the habitable zone, and that the system hosts a remarkable co-orbital gas giant configuration. We show that the existence of these three additional planets are readily amenable to testing with available data from both ground-based radial velocity data and space-based TESS photometry. A periodogram search of the available radial velocities show no compelling signals at the claimed periods, and the TESS photometry effectively rules out these planets assuming a transiting configuration. While it is doubtlessly possible that additional planets orbit GJ 3470, there is no evidence to date for their existence, and the available data conclusively rule out any planets similar to those considered in this text.

\end{abstract}


\section{Introduction}
\label{I}

The wonder of the heavens has captivated human imagination since before the advent of written language. Many people without formal training and access to the most advanced telescopes have desired to be a part of the search for discoveries, and new methods over the past few decades in crowd-sourcing have resulted in the modern "citizen scientist." Projects like Galaxy Zoo and Backyard Worlds: Planet 9 have allowed science enthusiasts to participate constructively in the search for knowledge. 

With the past thirty years bringing a bountiful harvest of extrasolar planet discoveries, the search for planets around nearby stars has become a particularly interesting area of research for both the public and the astronomical community. The increasing availability of knowledge, skills and talent among the general public have allowed for a diversification of the contributions of citizen scientists beyond the rather limited, structured environments like those aforementioned. Now anyone can search public data and make discoveries. 

Ultimately, the limited structured environments do have benefits, however. They allow claimed discoveries to go through a more rigorous process of vetting before being pushed to the forefront of public knowledge.

GJ 3470 is a nearby ($\pi$ $\sim$ 30 mas) bright ($J$ $\sim$ 8.8) M dwarf that hosts a transiting hot Neptune-like planet \citep{Bonfils2012}. The planet is of considerable interest for followup transmission spectroscopy with JWST \citep{Benneke2019}, so learning about its planetary system would offer a better context in which to understand it. In 2020, some ground-based photometry conducted by some citizen scientists ~\citep{TheExoplanetsChannel} was gathered and used to support a claim that a Saturn-sized planet (GJ 3470 c) had been detected in transit with an orbital period of 66 days ~\citep{Scott2020}. More recently, in 2023 the same group presented the reported discovery of two additional Jovian planets around the star (GJ 3470 d and e) in a co-orbital configuration with period of $\sim$15 days ~\citep{Scott2023}, while revising the period of the "c" candidate to $\sim$33 days.

Co-orbital planet pairs are expected to be detectable in transit, as well as through Doppler spectroscopy (~\cite{Laughlin_2002}, \cite{Leleu2017}). Though while examples of co-orbital celestial bodies are known to exist -- the trojan asteroids of the giant outer solar system planets and some of the moons of Saturn -- to date, no co-orbital extrasolar planet pairs have been confidently identified, though a possible Doppler detection of a co-orbital configuration at LHS 1140 was reported by \citep{Lillo-Box2020}. Consequently, the confirmation of a co-orbital planet pair would constitute a significant discovery, and we must therefore be cautious in advance of proposing such a claim.

The discovery of the known transiting hot Neptune around GJ 3740 was the result of a ground-based Doppler spectroscopy campaign that ultimately extended a decade. One might naturally wonder, therefore, if the available radial velocities can be used to investigate the presence of multiple additional giant planets around this star. Additionally, the Transiting Exoplanet Survey Satellite (hereafter TESS) mission obtained photometric coverage of the star in late 2021. In this work, we show that the existence of additional planets claimed by \citep{TheExoplanetsChannel} are ruled out by the available data.

\section{Doppler Spectroscopy}

We downloaded the 125 public radial velocity observations available from the ESO data access website from the HARPS spectrometer and analyzed them with Systemic \citep{Meschiari2009}. These radial velocities span almost ten years from 07 December 2008 through 09 April 2018. 

A Lomb-Scargle periodogram of the velocities produces two peaks: one at $3.3366$ days and another at $1.4224$ days. After fitting the data with a planet of the period known from \cite{Awiphan2016} of $3.3366496$ days, both peaks disappear, demonstrating the $1.4224$ day peak is an alias of the $3.3366$ day peak, which corresponds to the known transiting hot Neptune GJ 3470 b.

After accounting for the known planet, the periodogram of the residuals shows no significant peak. The strongest peak is at 383 days, with a $FAP=21\%$. As such, there is no evidence in the HARPS data for additional planets orbiting GJ 3470.

Would we have identified additional planets similar to the ones suggested by \cite{Scott2020} and \cite{Scott2023}? We can calculate the radial velocity semi-amplitude $K_b$ of a planet with a known orbital period $P$ and mass $M_p$ with

\[
K=\left(\frac{2\pi G}{P}\right)^{1/3}\frac{M_p \sin i}{(M_*+M_p)^{2/3}}
\]

\cite{Stefansson2022} analyzed radial velocities from HARPS, HIRES and HPF and found $K_b = 8.03^{+0.38}_{-0.37}$ ms$^{-1}$. This is consistent with our measured $K$ of $9.08$ though much more precise (Systemic does not permit fitting $K$ as a probability distribution from which to derive $\sigma_K$). Injecting a signal into the radial velocity data with a sufficiently high amplitude $K$ produces an \textit{anti}-signal in the residuals and hence in the periodogram. Anti-signals that produce periodogram peaks with a FAP $< 0.01$ are taken to define the limit of our ability to rule out additional planets, because any signal with a lower amplitude $K$ (and hence higher false alarm probability) would have been missed by this analysis with a probability of $>1\%$.

For the period of the proposed c planet, at 33 days, we find that a signal of $K\sim4.27$ ms$^{-1}$ would have been detectable with a probability of over 99\%. This corresponds to a planet with a minimum mass of $M_c \sin i \sim 0.044 M_J$, slightly less massive than the known (and confirmed) planet b. 

For the periods of the proposed d and e planets, at $\sim$15 days, we find that a signal of $K\sim3.49$ ms$^{-1}$ would have been detectable with a probability of over 99\%. This corresponds to a planet with a minimum mass of $M_{d,e}\sin i\sim0.027 M_J$.

\begin{figure}
    \centering
    \includegraphics[width=90mm,scale=0.5]{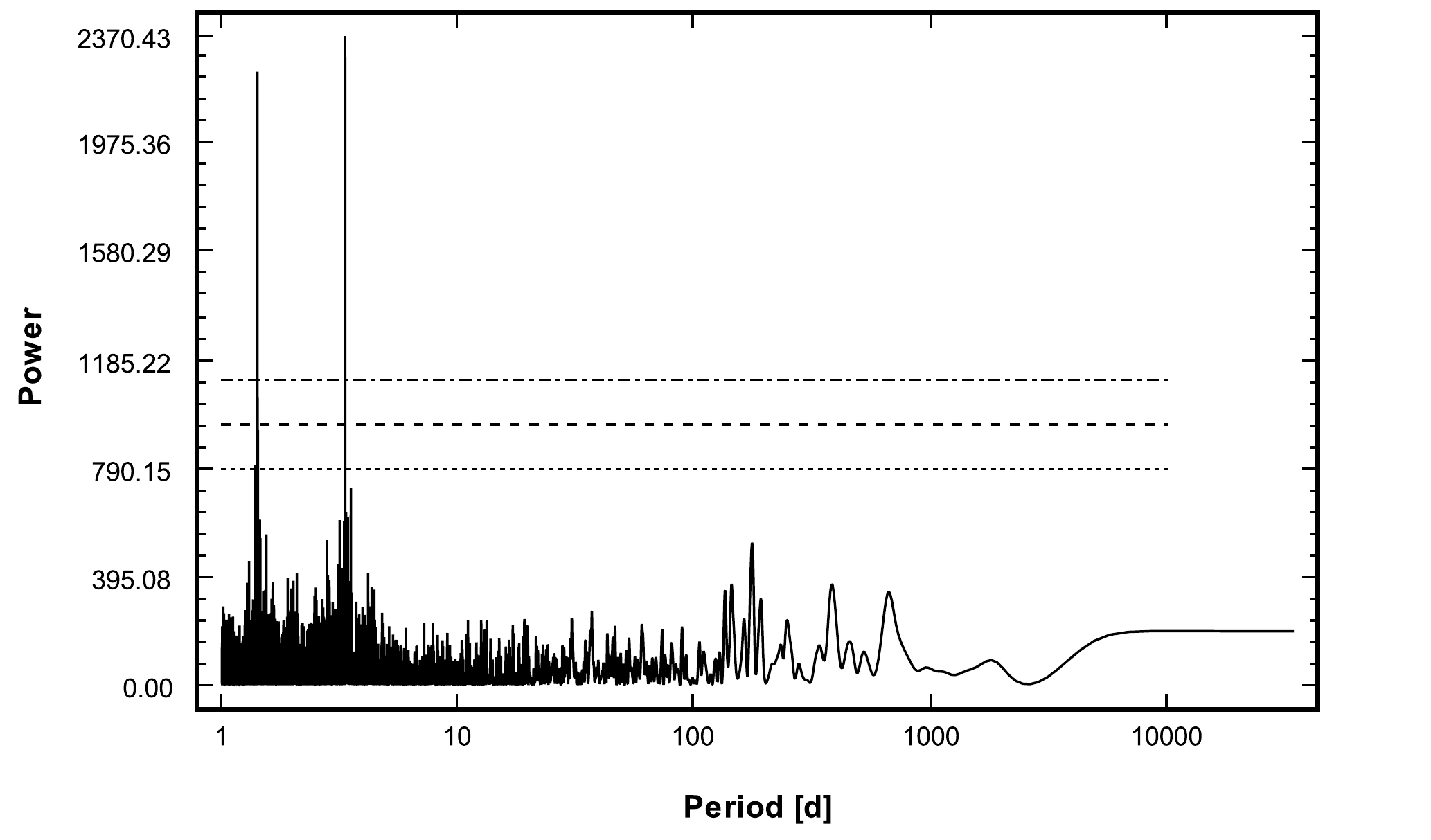}
    \caption{Lomb-Scargle Periodogram of the HARPS radial velocities for GJ 3470. The known transiting hot Neptune GJ 3470 b is readily apparent.}
    \label{fig:GJ3470 Periodogram}
\end{figure}

Given the radius given for the c planet in \citep{Scott2020} of 58,725 km, and the radii given for the d and e planets in \citep{Scott2023}, we find densities for all three proposed planet candidates of $<0.098$ g cm$^{-3}$, $<0.016$ g cm$^{-3}$ and $<0.076$ g cm$^{-3}$ for c, d and e, respectively. These densities, which should be considered to be conservative upper limits, are strikingly low, but not \textit{impossible}. Indeed a class of "super-puff" planets with extremely low densities was revealed by the Kepler mission with densities down to $\sim0.01$ g cm$^{-3}$, and recent work has shown that circumplanetary rings can inflate a planet's transit depth, producing a low estimated density when not taken into account \citep{Akinsanmi2020}. One must ask, however, why these planets were not detected by ~\cite{Stefansson2022}, who carried out a far more detailed treatment of the radial velocity behaviour of the star.

To address the possibility, however remote, that super-puff planets could explain the transit-like signals observed by \cite{Scott2020} and \cite{Scott2023}, we turn to TESS photometry.

\section{TESS Photometry}

The TESS spacecraft observed GJ 3470 nearly continuously from 12 October 2021 to 30 December 2021, providing a valuable dataset to check the validity of these planets.

TESS light curves of GJ 3470 were downloaded using the tools provided by the Lightkurve python package \citep{Lightkurve}. Data from Sectors 44 through 46 were extracted from the resulting lightcurve files. The PDCSAP flux of the lightcurve was used for our analysis, since it is cleaned and free from systematic effects and other unwanted flux jumps and contamination. The three lightcurves were then combined and detrended using a biweight filter, all done using the python detrending package Wōtan \citep{Hippke2019}. The resulting lightcurve is nearly continuous over a baseline of 78 days, which is long enough to detect multiple transits of all the claimed planets. When choosing a window size for the biweight filter, we took in mind the duration of the longest claimed transit (about three hours) and chose a detrending window of about six times this length ($0.75$ days) in order to avoid detrending out any possible transits of the claimed planets. 

\begin{figure}
    \centering
    \includegraphics[width=90mm,scale=0.5]{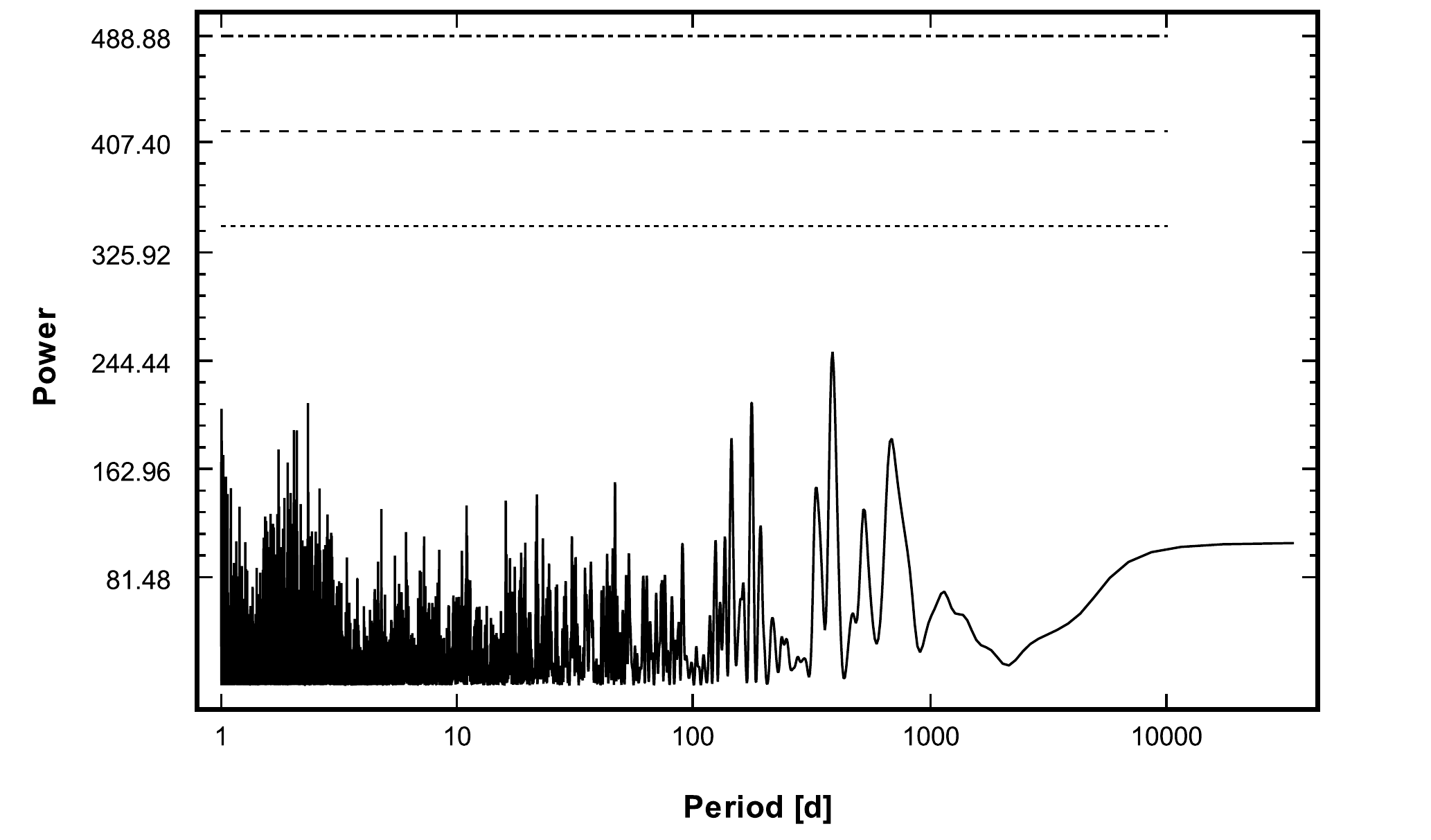}
    \caption{Lomb-Scargle Periodogram of the residuals for the HARPS radial velocities for GJ 3470 after the one-planet fit. No signals are apparent, which immediately casts doubt on the validity of the claimed exoplanets.}
    \label{fig:GJ3470 Periodogram of Residuals}
\end{figure}

 The next step in the processing is the removal of the transits of GJ 3470 b, to ensure that they do not interfere with the search for transits of additional planets. The data was phase-folded across the period of GJ 3470 b, and then using information from this phase-folded data, the transits were removed from the original unfolded light curve. To do this, we adopt the planetary orbital period from \cite{Awiphan2016} of $3.3366496$ days.

The data was then phase-folded to the orbital period of each of the three claimed planets, with orbital period values adopted from \cite{Scott2023} and a visual inspection of the phase-folded lightcurves was done to examine for possible transits. 

While the phase coverage across the phase-folded lightcurve of GJ 3470 at planet c's orbital period initially appears to be complete, a closer inspection reveals a 0.8 day long gap in the light curve centered at phase 0.417. Since this gap is longer in duration than the claimed transit, we must confirm that the transit does not fall in this gap in order to conclusively refute or confirm possible transits of c. When accounting for the detection of the ingress of the transit, the available window in the gap in the lightcurve in which we could have missed a transit of the purported GJ 3470 c decreases to about 0.75 days. If the transit of GJ 3470 c were to be at a random location in the phase-folded light curve, then by comparing the areas of the gap in the lightcurve to the area of the lightcurve as a whole, the chance of the transit falling in the gap is only around $\sim 2.3\%$. In any case, we took the mid-transit time of the last transit and calculated the time of the first transit in the TESS lightcurves, and found that it should occur near phase 0 in our phase-folded lightcurve. This area does not fall within the previously mentioned gap. As a result, we can effectively ignore this part of the light curve as it does not align with the predicted time of transit.

In the case of planets d and e, which are claimed to follow a co-orbital configuration, gravitational interactions between the two planet are expected to cause changes in the orbital period and mid-transit times, rendering the values obtained in \citep{Scott2023} unreliable for our TESS data. Fortunately for these two planets, complete phase coverage was achieved for the period-folded light curve, and therefore the refutation or confirmation of any potential planets becomes much simpler, with no need to check for the mid transit time of the planet, since we can rule out transits across the entirety of the orbital period.

\begin{figure}
    \centering
    \includegraphics[width=90mm,scale=0.5]{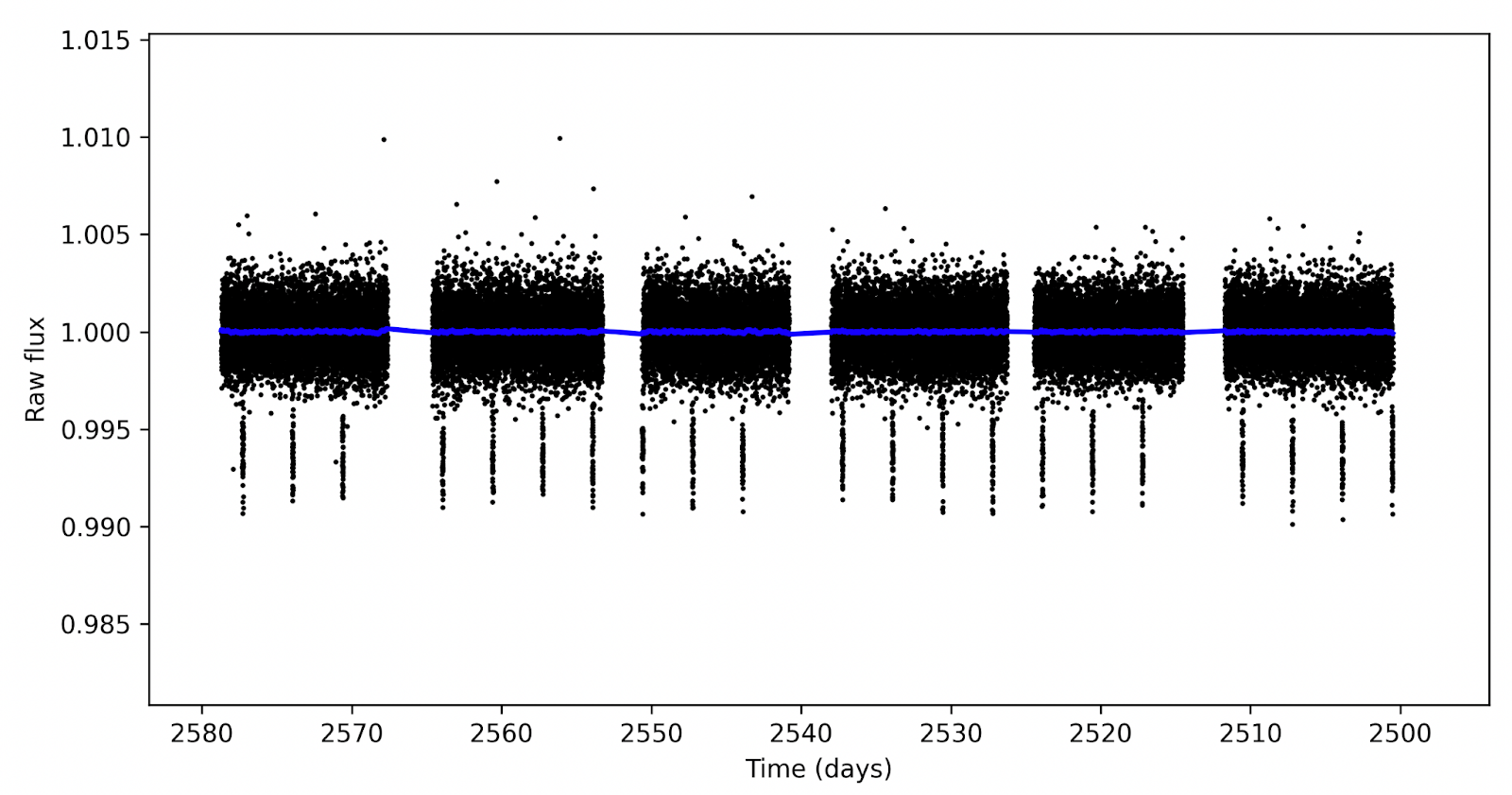}
    \caption{Detrended TESS lightcurve of GJ 3470. Aside from the transits of GJ 3470 b, no transits with depths greater than $\sim 0.3\%$ are seen in the lightcurve, which immediately casts doubt on the validity of the claimed exoplanets.}
    \label{fig:my_label}
\end{figure}

In the case of planet candidate c, we do not detect any transits deeper than the estimated 3$\sigma$ photometric error. We estimate this value by binning the phase-folded lightcurve into units of the length of the transit, and then calculating the photometric error by finding the standard deviation of the points inside each bin divided by the square root of the number of points in the bin. The resulting value was $0.14\%$, much deeper than the claimed transit depth of $0.84\%$, conclusively refuting the claimed planet GJ 3470 c. 

In the case of planet d, using the same calculation method, we can rule out transits deeper than $0.04\%$, much deeper than the claimed transit depth of $1.4\%$, and as a result we can conclusively refute the planet candidate GJ 3470 d. In the case of planet e, we can rule out transits deeper than $0.04\%$, much deeper than the claimed transit depth of $0.5\%$, and as a result we conclusively refute the planet candidate GJ 3470 e. We have, therefore, shown that the TESS photometry conclusively disproves the existence of all three planets presented by \cite{Scott2023}.

\section{Conclusions}

The available radial velocity data shows that GJ 3470 is bereft of short-period giant planets beyond the known transiting hot Neptune-like planet GJ 3470 b that are consistent with the claims by \cite{Scott2023}. The Doppler non-detection of the proposed transiting Jovian planets hints at unphysical densities for these planets. Nevertheless, in the event that the planets are truly novel objects, we examine TESS photometry to evaluate whether or not any evidence of their existence can be found there and find that it demonstrates, conclusively, that there are no transiting planets consistent with the claims of \cite{Scott2023}. We conclude that the apparent transit signals in the ground-based data are mostly likely noise or systematics
in the lightcurves.

The existence of a co-orbital pair, we would think, would inspire a more rigid attempt to validate the existence of these planets, since no such example has been confirmed to date. Additionally, \cite{Stefansson2022} found a polar orbit ($\lambda=98^{+15}_{-12}$ degrees) for the genuine planet GJ 3470 b from the Rossiter-McLaughlin effect in the measured radial velocities. This hints at a dynamically excited evolution of the system, which would make the existence of the proposed arrangement rather unexpected, though multi-planet systems have been identified in misaligned configurations before \citep{Huber2013}.

This work highlights the need to make use of available data resources to verify claims of new discoveries by citizen scientists. The data is often free and available. Anyone can use it. A sort of "due-diligence" should be seen to exist to apply available high-quality data taken by telescopes and institutions out of reach for citizen scientists that may conclusively test these hypotheses. Citizen scientists still have a role to play in astronomy, even outside the constraints of dedicated crowd-sourcing projects. We humbly present this work, itself drafted by citizen scientists without formal training, as an example, and we hope it inspires other citizen scientists to continue to do high quality work that makes valuable contributions to the field.

\acknowledgements


\bibliography{refs.bib}
\bibliographystyle{plainnat}

\clearpage

\begin{figure}
    \centering
    \includegraphics[width=190mm,scale=0.5]{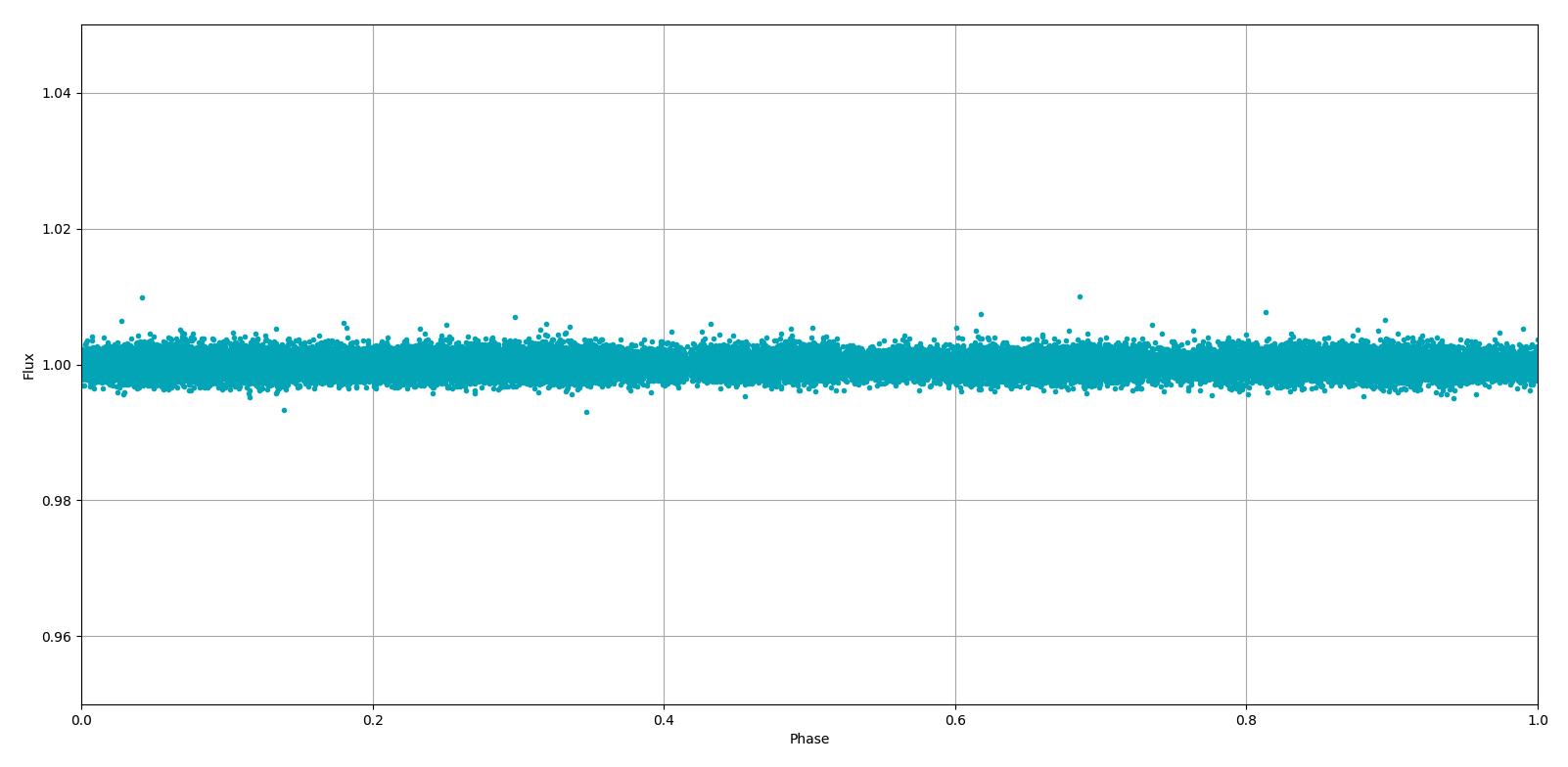}
    \caption{TESS Lightcurve of GJ 3470 phase-folded to the 33-day period of planet candidate c, with transits of b removed. The small 0.8-day gap in the data is inconspicuous at this scale and is not visible.}
    \label{fig:TESS GJ 3470 33 days}
\end{figure}

\begin{figure}
    \centering
    \includegraphics[width=190mm,scale=0.5]{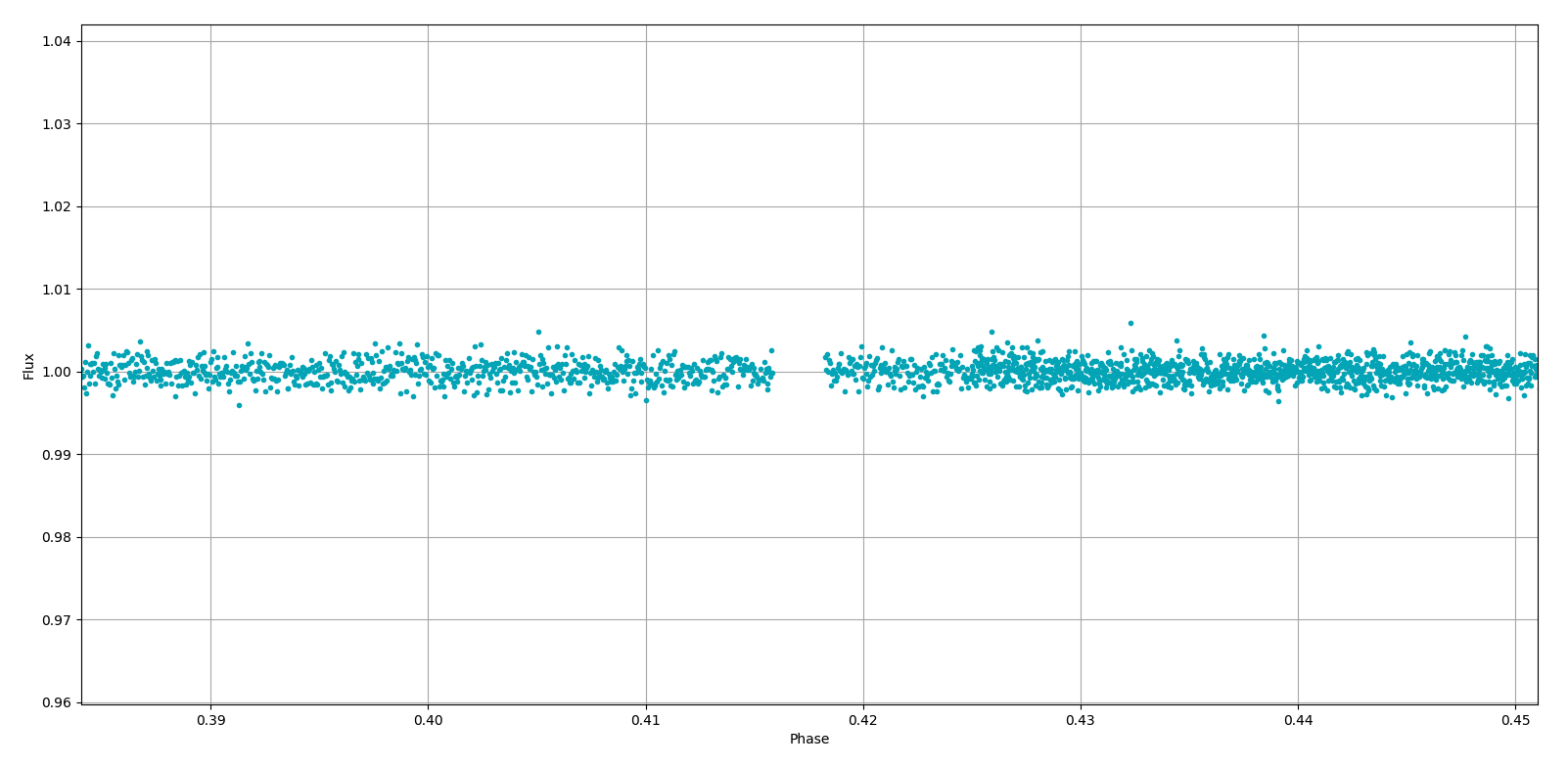}
    \caption{Close-up of the previous figure, showing the 0.8-day gap.}
    \label{fig:TESS GJ 3470 0.8-day gap}
\end{figure}

\begin{figure}
    \centering
    \includegraphics[width=190mm,scale=0.5]{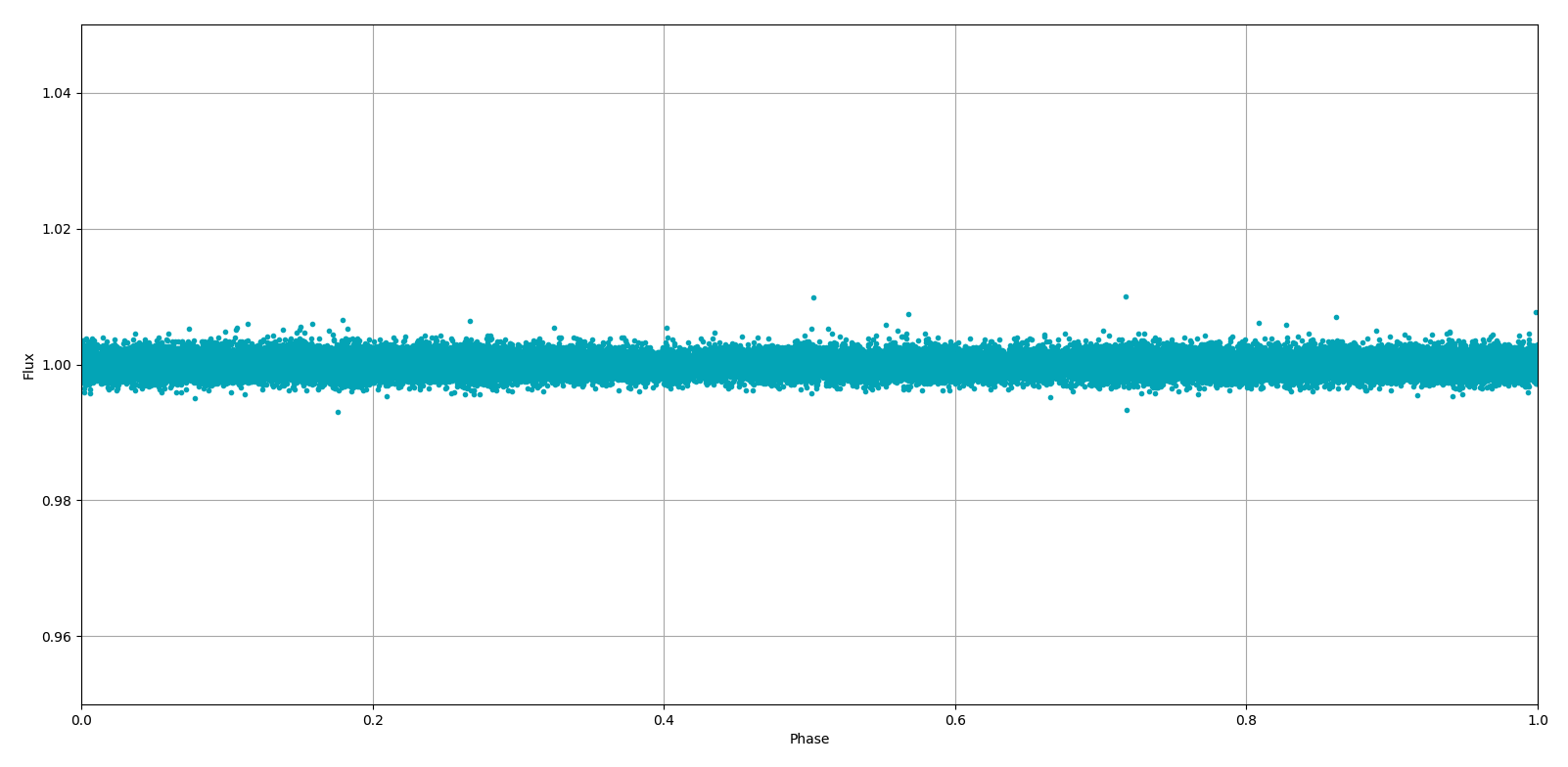}
    \caption{TESS Lightcurve of GJ 3470 phase folded to the 15-day period of planet candidate d, with transits of planet b removed.}
    \label{fig:TESS GJ3470 15 days}
\end{figure}

\begin{figure}
    \centering
    \includegraphics[width=190mm,scale=0.5]{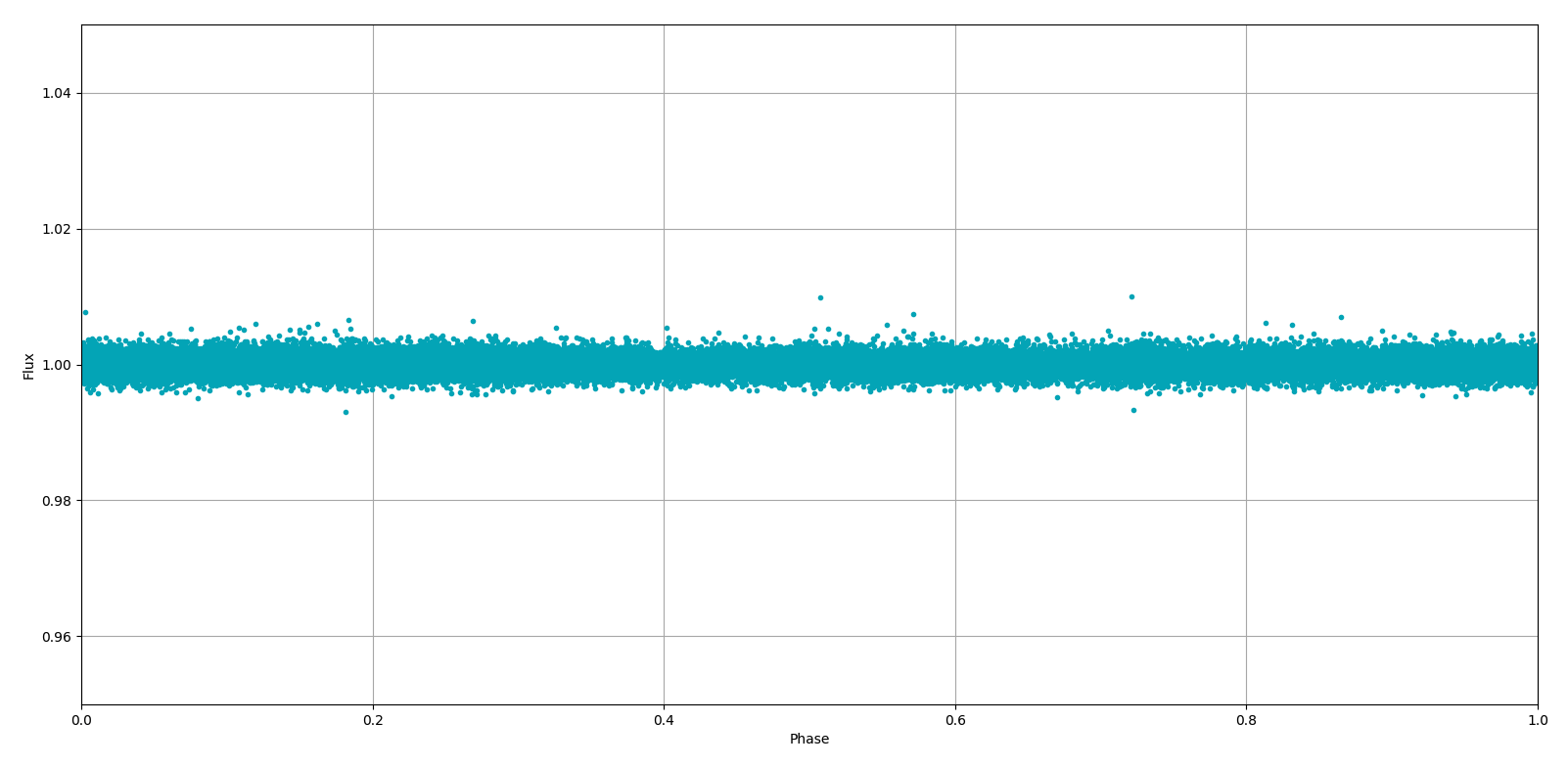}
    \caption{TESS Lightcurve of GJ 3470 phase-folded to the 15-day period of planet candidate e, with transits of planet b removed. The lightcurve appears almost identical to that of the previous figure because of the similar orbital periods of the planet candidates d and e. }
    \label{fig:TESS GJ 3470 15days2}
\end{figure}

\end{document}